# A Pipeline for ADNI Resting-State Functional MRI Processing and Quality Control

Saige Rutherford[1], Zeshawn Zahid[1], Robert C. Welsh[2], Andrea Avena-Koenigsberger[3,4], Dustin B. Hammers[5], Vincent Koppelmans[6,7], Alzheimer's Disease Neuroimaging Initiative[*], Amanda F. Mejia[1+]

[1]Department of Statistics, Indiana University, Bloomington, USA
[2]Psychiatry and Biobehavioral Sciences, University of California, Los Angeles, USA
[3]Indiana Alzheimer's Disease Research Center (IADRC), Department of Radiology and Imaging Sciences, Indiana University School of Medicine
[4]Center for Neuroimaging (CfN), Medical Imaging Research Institute (MIRI), Department of Radiology and Imaging Sciences, Indiana University School of Medicine
[5]Department of Neurology, Indiana University School of Medicine, Indianapolis, USA
[6]Department of Psychiatry, University of Utah, Salt Lake City, USA; [7]Huntsman Mental Health Institute, University of Utah, Salt Lake City, USA
[+]Corresponding author: Saige Rutherford
Email: saige.rutherford@gmail.com
Address: 729 E. 3rd Street, Bloomington, IN 47405

**Abstract**

**INTRODUCTION:** The Alzheimer's Disease Neuroimaging Initiative (ADNI) provides a comprehensive multimodal neuroimaging resource for studying aging and Alzheimer's disease (AD). However, its resting-state functional MRI (rs-fMRI) data remain underutilized, in part due to challenges with data processing driven by considerable variability in acquisition protocols, inconsistent organization, missing or incomplete imaging sessions, and frequency of data quality issues due to participant motion and aging-related factors.

**METHODS:** To facilitate widespread use of ADNI rs-fMRI data, we developed an open-source pipeline encompassing the download and organization of imaging and clinical data, conversion to Brain Imaging Data Structure (BIDS), preprocessing with fMRIPrep, and automated quality control (QC). Our QC considers acquisition parameters and 16 structural and functional MRIQC image-quality metrics, flagging scans that fall outside acceptable ranges based on expert input and robust outlier detection techniques. Processed imaging sessions are matched to temporally aligned clinical assessments to produce an analysis-ready dataset.

**RESULTS:** Applying the pipeline to more than 2,800 sessions from ADNI GO, ADNI 2, and ADNI 3 yielded 2,086 quality-checked functional and structural MRI sessions from 935 unique participants. Exclusions occurred at every processing stage and were driven primarily by BIDS conversion errors, image quality flags, and preprocessing- or motion-related failures.

**DISCUSSION:** The open-source pipeline described in this work produces one of the largest pre-processed, quality controlled, and analysis-ready functional and structural MRI datasets yet available for studying aging and AD. It lowers barriers to large-cohort and longitudinal neuroimaging research and supports large-scale neuroimaging biomarker discovery in AD.

---

[*] Data used in preparation of this article were obtained from the Alzheimer's Disease Neuroimaging Initiative (ADNI) database (adni.loni.usc.edu). As such, the investigators within the ADNI contributed to the design and implementation of ADNI and/or provided data but did not participate in analysis or writing of this report. A complete listing of ADNI investigators can be found at: http://adni.loni.usc.edu/wp-content/uploads/how_to_apply/ADNI_Acknowledgement_List.pdf

# 1 Introduction

The Alzheimer's Disease Neuroimaging Initiative (ADNI) is a large multicenter consortium spanning more than 60 academic and clinical sites across the United States and Canada. ADNI collects longitudinal MRI, PET, clinical, cognitive, and biochemical data from cognitively unimpaired older adults, individuals with mild cognitive impairment (MCI), and patients with Alzheimer's disease (AD)[1], with the goal of standardizing biomarkers and outcome measures for AD clinical trials and developing an accessible longitudinal data repository[2]. ADNI has collected data continuously since 2004 across several phases (Table 1).

**Table 1**. Overview of the phases of ADNI and the primary goal, timeline/duration, funding, and cohort description of each phase. Each cohort after ADNI 1 includes the previous cohort(s). This table was recreated from ADNI's website. CU = Cognitively Unimpaired, MCI = Mild Cognitive Impairment, AD = Alzheimer's Disease, DEM = Dementia.

| | **ADNI 1** | **ADNI GO** | **ADNI 2** | **ADNI 3** | **ADNI 4** |
|---|---|---|---|---|---|
| **Timeline (Duration)** | 2004-2010<br><br>(5 years) | 2009-2011<br><br>(2 years) | 2011-2016<br><br>(5 years) | 2016-2022<br><br>(5 years) | 2022-2027<br><br>(5 years) |
| **Primary Goal** | To develop biomarkers as outcome measures for clinical trials. | Examine biomarkers in earlier stages of disease. | Develop biomarkers as predictors of cognitive decline, and as outcome measures. | Study the use of tau PET and functional imaging techniques in clinical trials. | Improve generalizability of AD research. |
| **Cohort** | 200 CU, 400 MCI, 200 AD | +200 early MCI | + 150 CU, 150 elderly MCI, 150 late MCI, 150 AD | + 133 CU, 151 MCI, 87 AD | + 200 CU, 200 MCI, 100 AD/DEM |

ADNI has been instrumental in advancing understanding of neurodegeneration by providing large-scale, longitudinal neuroimaging and biomarker data across the spectrum of cognitive aging and dementia[3–6], with more than 6,000 publications, 30,000 data use agreements, and 405 million image downloads to date[7], including 752 papers in 2025 alone. In contrast, relatively few studies have leveraged ADNI resting-state fMRI (rs-fMRI): a targeted PubMed search (full search strategy in Supplementary Material) identified only 352 publications using ADNI rs-fMRI as of 2025, just 5.9% of all ADNI publications. In 2025 alone, only 48 of 752 ADNI papers (6.4%) used rs-fMRI (Figure 1).

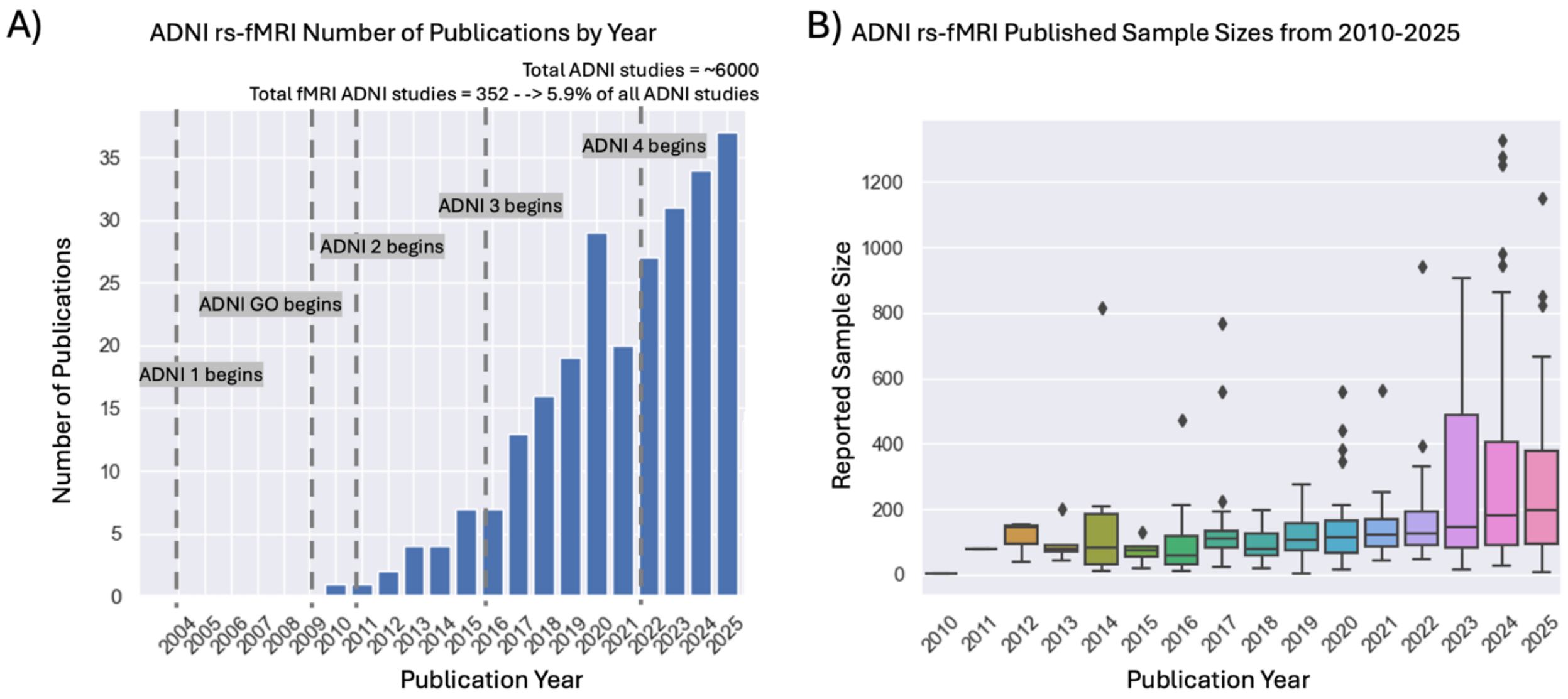

**Figure 1. A)** The number of published papers using ADNI rs-fMRI data per year (2010-2025). A full list of these papers is provided in the supplement (Supplemental File 1). **B)** The reported sample sizes in each publication using ADNI rs-fMRI data, per year (2010-2025). The average sample size across all publications is 195 subjects (SD=222).

This gap is not fully explained by the later start of rs-fMRI acquisition (ADNI-GO in 2009 vs. ADNI-1 in 2004); instead, it likely reflects structural barriers to reuse rather than a lack of scientific interest. Rs-fMRI data are inherently more complex, require more extensive preprocessing, and are sensitive to acquisition and quality-control variability, factors reflected in the wide variation in published sample sizes, from 2 to 1326 subjects (Figure 1). This underutilization represents a major opportunity: ADNI rs-fMRI is a rich, longitudinal, yet largely untapped resource for studying functional brain changes in aging and AD, one that modern data standards and processing tools are positioned to unlock.

A reproducible, transparent, semi-automated preprocessing pipeline for ADNI functional and structural MRI data will accelerate large-scale secondary analyses of this resource and provide the transparent documentation that reproducible neuroimaging research requires[8,9]. Here we describe a pipeline that converts ADNI functional and structural MRI data into the Brain Imaging Data Structure[10], applies standardized preprocessing, performs automated quality assessment, flags data requiring troubleshooting, and integrates clinical assessments. The pipeline combines established tools including Clinica[11,12], fMRIPrep[13], and MRIQC[14], with bespoke quality control tools and an optional containerized environment for reproducibility across computing systems. Our configuration files and processing scripts provide a complete blueprint for producing an analysis-ready dataset within the constraints of the ADNI data use agreement, supporting functional biomarker discovery in AD and related neurodegenerative conditions.

## 2 Methods

Here, we describe a complete data curation pipeline for ADNI functional and structural MRI, organized according to nine main workflow steps (Figure 2) implemented in a public GitHub repository (https://github.com/statMINDlab/ADNI_fMRI_protocol). For each step, we specify the data inputs, processing operations, software tools and versions, quality control procedures, and known issues specific to ADNI. All scripts and supporting files referenced below are available in the repository. This pipeline currently supports the ADNI GO, ADNI 2, and ADNI 3 datasets.

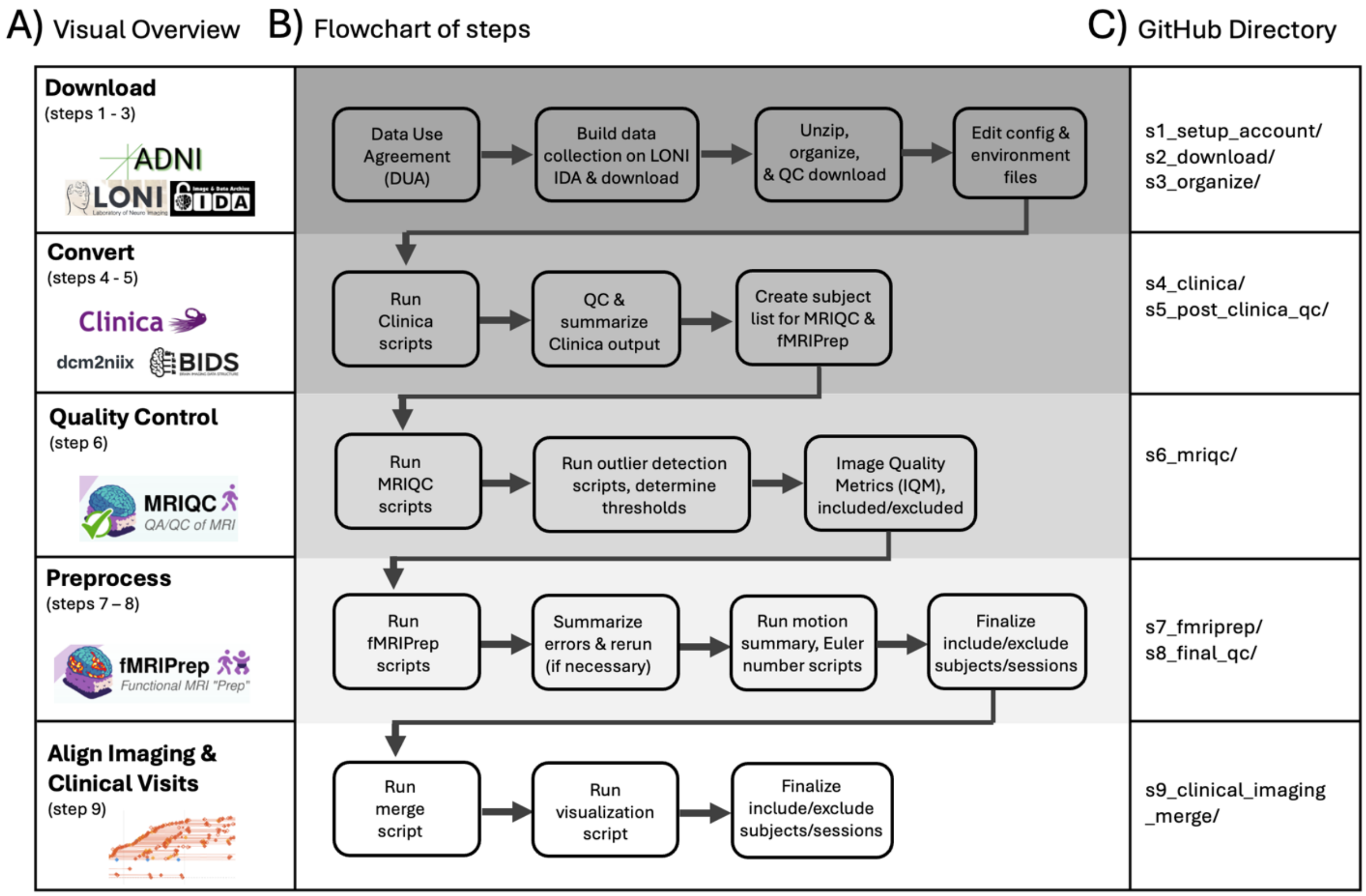

**Figure 2.** Overview of the nine steps in the ADNI fMRI data curation and processing pipeline. Step numbers (panel A) correspond to the detailed steps in the GitHub repository (panel C) (https://github.com/STATMINDlab/ADNI_fMRI_protocol).

Data used in the preparation of this article were obtained from the ADNI database (adni.loni.usc.edu). ADNI was launched in 2003 as a public-private partnership, led by Principal Investigator Michael W. Weiner, MD. The primary goal of ADNI has been to test whether serial MRI, PET, other biological markers, and clinical and neuropsychological assessment can be combined to measure the progression of mild cognitive impairment (MCI) and early AD. For up-to-date information, see https://adni.loni.usc.edu/.

*Step 1: Account Setup and Data Use Agreement*

Raw imaging and clinical data are accessed through the Laboratory of Neuro Imaging Image and Data Archive (LONI/IDA) (https://ida.loni.usc.edu/) after account registration and approval of ADNI's Data Use Agreement (DUA). The protocols here provide detailed instructions (further detailed instructions are available in the folder s1_setup_account/) for users to reproduce the entire workflow independently; however, no raw data are shared via our repository, in compliance with ADNI's DUA. All processing is performed within secure institutional or high-performance computing (HPC) environments, consistent with ADNI DUA requirements.

*Step 2: Data Selection and Download from LONI*

The second step involves selecting eligible ADNI sessions and downloading the associated imaging files, metadata spreadsheets, and study information packages (further instructions in s2_download/). Using the "Image Collections" tool in LONI/IDA, users can filter for resting-state fMRI sequences and associated structural sequences (T1 & T2) across the ADNI GO, ADNI 2, and ADNI 3 phases. ADNI 1 is excluded because it does not include resting-state fMRI, and Clinica does not yet support ADNI 4. The GitHub repository for the pipeline includes step-by-step screenshots and instructions for building reproducible LONI queries. All downloaded ZIP files, including raw imaging archives (DICOM format), metadata spreadsheets (e.g., ADNIMERGE.csv), and ADNI study files (e.g., clinical/behavioral data), should be stored as defined in the user's configuration file. For the current manuscript, data released prior to October 1$^{st}$, 2025, were downloaded and used for the development and testing of the pipeline.

*Step 3: DICOM Organization and Integrity Checks*

After download, all ZIP archives are uncompressed, reorganized into a standardized directory structure, and subjected to basic data quality checks. Scripts in s3_organize/ verify the DICOM download (expected number of directories based on image collection) and map the heterogeneous DICOM directory names to their corresponding imaging modalities (i.e., T1, T2, fMRI (i.e., BOLD)).

*Step 4: Imaging and Clinical Data Conversion Using Clinica*

Clinica is an open-source software platform that provides standardized converters from raw acquisition formats (e.g., ADNI DICOMs) to the Brain Imaging Data Structure (BIDS). DICOM files are converted into BIDS format (v1.10.1)[10] format using Clinica (v0.10.1)[11,12]. The command-line interface and configuration parameters are provided in s4_clinica/. Clinica serves two main functions: i) DICOM to NIfTI conversion and metadata extraction (via dcm2niix) and ii) organizing data into BIDS format. Clinica cross-references session dates using several ADNI spreadsheets (e.g., ADNIMERGE, MPRAGEMETA, and quality control tables) to ensure accurate mapping of T1, T2, and BOLD series to visit codes. After conversion, the resulting dataset is validated using the BIDS Validator (v1.14.7). Each participant's BIDS dataset contains both structural (T1 and, if available, T2) and functional (BOLD) images, along with all required metadata fields necessary for downstream processing with fMRIPrep and MRIQC.

During the development of this pipeline, we identified several issues in the ADNI DICOMs that must be addressed prior to preprocessing. These include items flagged by Clinica (interslice distance variation and failed image conversion, both documented in s4_clinica/known_clinica_DICOM_errors.csv) and several problems we encountered that were not reported by Clinica (missing or incorrect PhaseEncodingDirection parameter, missing JSON sidecar files for T1 files, T1 DICOM conversion errors, and absent SliceTiming metadata for BOLD series acquired on Philips scanners). Our scripts automatically flag these issues. Subjects with unrecoverable errors are excluded at this stage. For a more detailed description of Clinica, including some of the decisions about which scan to select (e.g., if there are multiple T1 per session), we refer readers to their detailed documentation (https://aramislab.paris.inria.fr/clinica/docs/public/dev/Converters/ADNI2BIDS/) and user support Google group (https://groups.google.com/g/clinica-user).

*Step 5: Post-Clinica Metadata Correction and Data Quality Checks*

Philips BOLD scans lack the slice-acquisition timing metadata needed for slice timing correction during preprocessing with fMRIPrep. As such, Philips scans would be processed differently from scans acquired on Siemens and GE systems. To ensure consistent preprocessing across scanner manufacturers, we reconstruct the slice-acquisition timing for Philips scans using information available in tables downloaded from LONI (MAYOADIRL_MRI_FMRI_NFQ_*.csv). We then add it to the BOLD metadata prior to preprocessing to ensure slice timing correction is performed correctly. Scans for which slice timing cannot be determined unambiguously are flagged and excluded.

After DICOM and BIDS conversion, additional QC steps ensure that each session meets minimum functional and structural imaging criteria (s5_post_clinica_qc/). Five heuristics regarding the scanner parameters are applied (Table 2). These heuristics were determined by an expert in MR physics (author RCW) and were chosen to detect errors in the data that occurred during collection. All scanner parameters were reviewed across sites/scanners to aid in the heuristic threshold choices. A notebook (s5_post_clinica_qc/analysis/create_report/main.ipynb, also available as a Python script) creates per-site and per-manufacturer (i.e., Siemens, GE, Philips) QC plots, as well as the summary tables of scanner parameters and inclusion/exclusion sample sizes at each decision point. It also summarizes and categorizes Clinica output errors (e.g., missing or incorrect *PhaseEncodingDirection*, missing JSON sidecar files for T1 files, or T1 DICOM conversion errors). The final step of this notebook is to export a CSV file with the subjects and sessions to be passed on to MRIQC and fMRIPrep.

**Table 2.** Heuristics for excluding subjects/sessions based on scanner parameters.

| Metric | Exclusion Rule | Rationale |
|---|---|---|
| ScanDepth = $dim_3$ x $pixdim_3$ | Outside range [155, 180] | Ensures adequate brain coverage |
| Repetition Time (TR) | Outside [0.5–1.0] or [2.9–3.1] seconds | Detects mislabeled scans or protocol inconsistencies |
| Minimum total scan duration | < 300 seconds | Ensures ≥5 minutes of fMRI data |
| PercentPhaseFOV | < 72 | Indicates spatial under sampling |
| CoilString | Q-Body or BODY coils | Nonstandard or problematic coils |

*Step 6: Automated QC with MRIQC*

MRIQC is an open-source, automated quality control tool that computes standardized image quality metrics (IQMs) directly from T1-weighted and BOLD MRI data, without requiring a reference dataset[14]. Sixteen automated participant-level IQMs are generated by MRIQC (version 24.0.2) for each individual session, and group-level distributions are used to determine quantitative exclusion criteria. The full list and description of IQMs is provided in Table 3.

**Table 3.** Automated image quality metrics from MRIQC.

| Name | Used for | Description | Direction |
|---|---|---|---|
| Contrast to Noise Ratio (CNR) [17] | T1 | Extension of the SNR to evaluate how separated the tissue distributions of GM and WM are. | Higher = better |
| Signal to Noise Ratio (SNR) [18] | T1 | The mean intensity within gray matter divided by the standard deviation of the values outside the brain. | Higher = better |
| Tissue probability overlap map (TPM) of Cerebrospinal Fluid (CSF) | T1 | Overlap estimated from the CSF image and the corresponding maps from the ICBM nonlinear-asymmetric 2009c template. | Higher = better |
| TPM of Gray Matter (GM) | T1 | Overlap estimated from the GM image and the corresponding maps from the ICBM nonlinear-asymmetric 2009c template. | Higher = better |
| TPM of White Matter (WM) | T1 | Overlap estimated from the WM image and the corresponding maps from the ICBM nonlinear-asymmetric 2009c template. | Higher = better |
| Coefficient of Joint Variation (CJV) [19] | T1 | CJV of GM and WM: objective function for the optimization of INU correction algorithms. Higher values are related to the presence of heavy head motion and large INU artifacts. | Lower = better |
| Entropy Focus Criterion (EFC) [20] | T1 | Uses the Shannon entropy of voxel intensities as an indication of ghosting and blurring induced by head motion. | Lower = better |
| Full Width Half Max (FWHM (mm)) [21] | T1 | The FWHM of the spatial distribution of the image intensity values in units of voxels. | Lower = better |
| Signal to Noise Ratio (SNR) [18] | BOLD | The mean intensity within gray matter divided by the standard deviation of the values outside the brain. | Higher = better |
| Temporal Signal to Noise Ratio (TSNR) | BOLD | Median value of the tSNR map (average BOLD signal / temporal standard deviation). | Higher = better |
| AFNI's mean quality index (AQI) [22] | BOLD | Mean quality index as computed by AFNI's 3dTqual for each volume, it is one minus the Speaman's rank correlation of that volume with the median volume. | Lower = better |
| AFNI's outlier ratio (AOR) [22] | BOLD | Mean fraction of outliers per fMRI volume as given by AFNI's 3dToutcount. | Lower = better |
| Entropy Focus Criterion (EFC) [20] | BOLD | Uses the Shannon entropy of voxel intensities as an indication of ghosting and blurring induced by head motion. | Lower = better |
| Framewise displacement (meanFD) [23] | BOLD | Expresses instantaneous head-motion. Rotational displacements are calculated as the displacement on the surface of a sphere of radius 50 mm. | Lower = better |
| Full Width Half Max (FWHM (mm)) [21] | BOLD | The FWHM of the spatial distribution of the image intensity values in units of voxels. | Lower = better |
| Temporal derivative of RMS variance over voxels (DVARS) [24] | BOLD | Indexes the rate of change of BOLD signal across the entire brain at each frame of data. | Lower = better |

There are no established thresholds for these IQMs, to our knowledge. To determine cutoffs for these IQMs in an automated fashion, we use a robust statistical outlier detection method developed by our lab[15]. For each IQM, the distribution across sessions is transformed to approximately follow a Normal distribution using a statistically robust, flexible transformation to eliminate skew and heavy or light tails. In the standard Normal distribution, values beyond ±4 are extremely unlikely (probability < 0.000063). Thus, after transformation, values above 4 or below -4 (depending on whether higher or lower values of the IQM are considered better, see

Table 3) are flagged. These cutoffs are translated back to the original scale of the data by applying the reverse transformation. The code implementing this procedure and for producing visualizations is located in s6_mriqc/.

*Step 7: fMRIPrep Preprocessing*

Image preprocessing is performed using fMRIPrep version 25.2.3[16], which includes surface reconstruction and projection via Freesurfer. For more details of the fMRIPrep pipeline, see fMRIPrep's workflows documentation (https://fmriprep.org/en/stable/) and citation boilerplate (Supplemental methods). The scripts for this step and further instructions are in the directory s7_fmriprep/.

We systematically parsed fMRIPrep logs to categorize failures (missing BOLD files, TemplateFlow cache/bind issues, insufficient memory, FreeSurfer recon-all failures, and fieldmap-less distortion correction failures). We designed upstream checks and conditional execution logic, which ensure that jobs are only submitted when all required inputs are present. Full details of each failure mode, and how they are resolved, are provided in the Supplementary methods.

*Step 8: Final QC and Derivation of the Analysis Sample*

The final QC stage comprises quality checking the subjects and sessions that successfully completed fMRIPrep and all previous steps. This includes summarizing the quality of the Freesurfer surface reconstruction and checking the functional runs for motion. Surface reconstruction quality is checked using the Euler number, a measure of the topological complexity of the surface, since higher topological complexity is considered to be indicative of problems with surface reconstruction. It has been shown to correlate highly with manual quality checking ratings[25] and is the main metric used in automated QC methods[26]. Since the distribution of the Euler number varies across sites, we adopt a simple approach that involves scaling and centering the distribution within site and removing samples in the tail of the distribution[27]. After censoring based on a framewise displacement (FD) threshold of 0.5 mm, sessions with more than 30% of frames censored or less than 5 minutes of low-motion data remaining are excluded. These thresholds can be adjusted by users as desired. A consolidated inclusion/exclusion table (with reasons coded per session) is produced in s8_final_qc/. The final analysis-ready dataset includes all subjects meeting QC criteria across every stage of this protocol (DICOM download, Clinica DICOM-to-NIfTI conversion and BIDS formatting, MRIQC, and fMRIPrep preprocessing).

*Step 9: Merging of clinical and imaging visits*

The final step of the pipeline involves matching imaging sessions to clinical assessments for each participant. Because imaging and clinical assessments are not always collected on the same calendar date, each BIDS session is paired with the clinical visit whose assessment date falls closest in time, subject to a maximum allowable gap. The default maximum is 180 days, but this can be adjusted by the user via the configuration file. Sessions without a clinical visit within this window are flagged, rather than silently dropped.

Session dates are extracted during BIDS conversion (Section 2.4), and visit codes (VISCODEs) are obtained from ADNIMERGE. This is not simple renaming: ADNI VISCODEs (e.g., m06, m12) do not always correspond one-to-one with imaging visit numbers, so the matching is performed explicitly on visit dates, rather than assumed from VISCODE labels alone. The resulting merged table links each participant_id/session_id pair to its aligned clinical, cognitive, and diagnostic variables (including diagnosis at that visit, MMSE, ADAS, MOCA, and CDR scores) and is written to clinical_data.tsv. Other behavioral and clinical variables can be included if desired, with minor modifications to the code. Scripts implementing this matching procedure are provided in s9_clinical_imaging_merge/.

## 3 Results

Applying this protocol to ADNI GO, ADNI 2, and ADNI 3 (data downloaded through October 1, 2025) yields a final analysis-ready dataset of 935 participants and 2,086 rs-fMRI sessions. Table 4 reports the number of subjects and sessions retained and dropped at each stage of the pipeline, from the initial download (1,153 subjects, 2,893 sessions) to final inclusion, based on our ADNI data download. This illustrates the cumulative effects of download issues, metadata inconsistencies, BIDS conversion failures, MRIQC-based outlier detection, fMRIPrep preprocessing errors, and final motion/Euler-number quality thresholds on the resulting analysis sample. Documentation for producing a finalized sample-size summary (i.e., Table 4) directly from the pipeline outputs is provided in the Supplementary material and the accompanying GitHub repository (see code availability below). In the initial QC phases (BIDS-convert via Clinica and Post-Clinica QC), subjects were excluded due to missing data or metadata (28 subjects/210 sessions) or incompatible scanner parameters (described in Table 2) (20 subjects/62 sessions).

**Table 4.** Inclusion and exclusion sample sizes at each step of the protocol. The Dropped column reports how many subjects (sub) and sessions (ses) were excluded from the previous step (row). The final column reports the sample size after exclusions up to each step. The sample consists of data from ADNI-GO, ADNI-2, and ADNI-3, the data releases used in our demonstration of the pipeline.

| Pipeline Stage | GitHub Folder | Description | Dropped (Sub/Ses) | Subjects/ Sessions |
|---|---|---|---|---|
| Unzip & organize DICOMs | s3_organize/ | Clean directory tree; check integrity | - | 1153/2893 |
| BIDS-convert via Clinica | s4_clinica/ | NIfTI + JSON + sidecars | 28/210 | 1125/2683 |
| Post-Clinica QC | s5_post_clinica_qc/ | TR/coverage/duration checks | 20/62 | 1105/2621 |
| MRIQC | s6_mriqc/ | All IQMs present; MRIQC ran successfully | 12/46 | 1093/2575 |
| Post-MRIQC QC | s6_mriqc/ | Robust outlier detection | 25/86 | 1068/2489 |
| fMRIPrep | s7_fmriprep/ | Volumetric, surface, grayordinates outputs | 121/375 | 947/2114 |
| Final QC | s8_final_qc/ | Euler number, motion, ≥5 min usable data | 12/28 | **935/2086** |

As described above (Step 6), automated outlier detection identified structural and functional scans with atypical MRIQC image quality metrics (IQMs). Figure 3 visualizes the distribution of the IQMs for structural and functional data, along with the thresholds for exclusion. A scan was classified as failed if *any* IQM exceeded its threshold. If the T1 image failed, the BOLD image was also excluded since the BOLD cannot be successfully preprocessed without a high-quality T1 image from the same session. A total of 46 sessions were dropped due to missing IQMs (i.e., MRIQC failed, likely due to poor quality data), and 86 sessions were dropped due to IQMs outside of the determined thresholds.

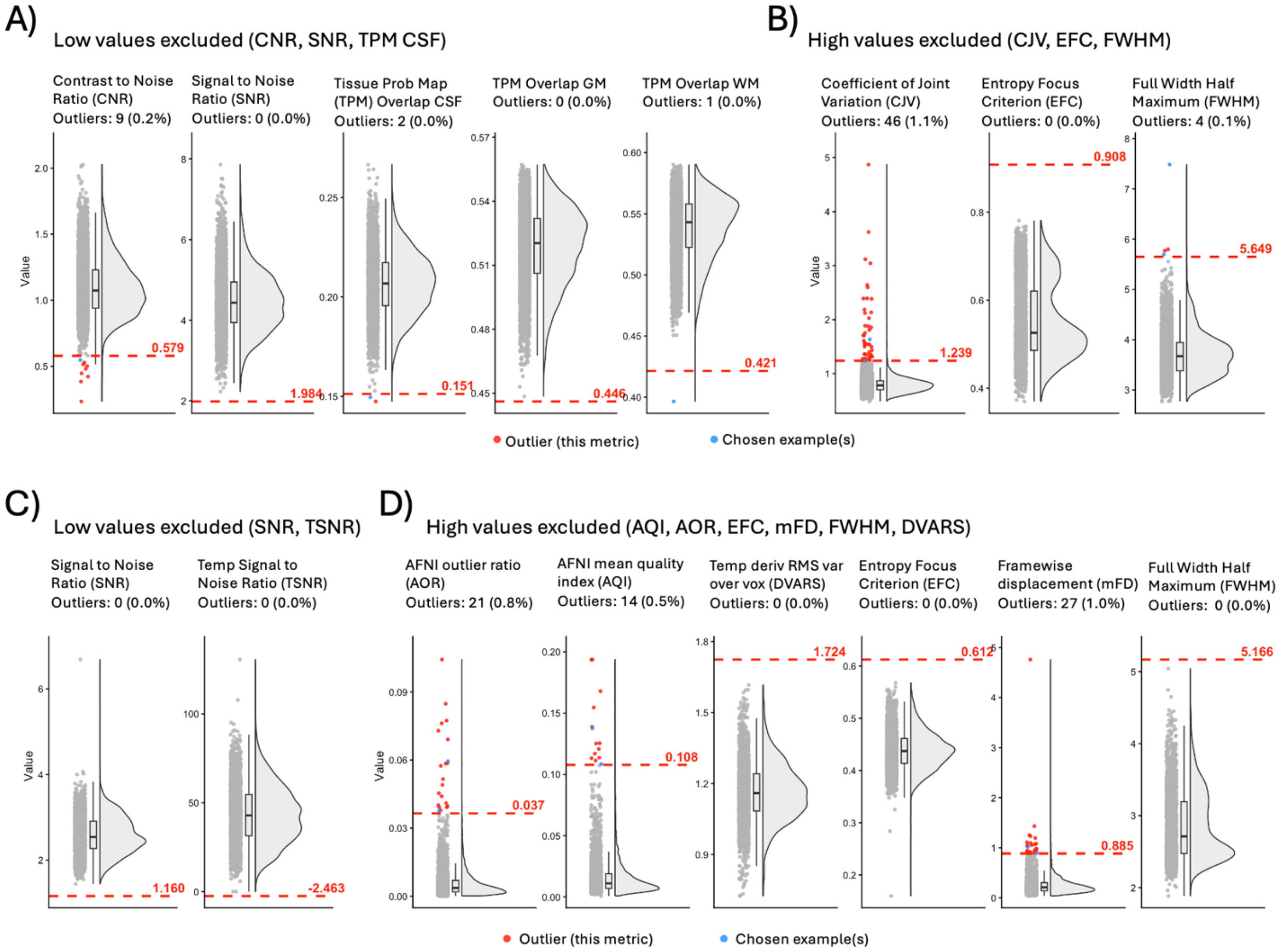

**Figure 3**. **Top row:** MRIQC Image Quality Metrics (IQMs) for structural (T1) images. These metrics are explained in Table 3. The red line indicates the outlier threshold. **A)** images below this line are excluded, and **B)** images above this line are excluded. Subjects were excluded if they had an outlier on any of these. **Bottom row:** MRIQC Image Quality Metrics (IQMs) for BOLD images. These metrics are explained in Table 3. The red line indicates the outlier threshold. **C)** images below this line are excluded, and **D)** images above this line are excluded. Sessions were excluded if they had an outlier on any of these.

Examples of scans excluded for each metric are shown in the supplement (Supplemental figures 1-7). Example scans were selected i) at random and ii) near the exclusion threshold to determine that the automatic metrics were catching true outliers with poor quality data. We also visualize the distribution of each IQM per site in the supplement (Supplemental figures 8-9). This approach offers an automated, reproducible, and interpretable strategy for identifying problematic scans in heterogeneous, multicenter datasets.

During fMRIPrep preprocessing, a small number of sessions required reprocessing without fMRIPrep's fieldmap-less distortion correction (predominantly Philips acquisitions missing *PhaseEncodingDirection* metadata). A total of 375 sessions (121 subjects) were excluded due to unrecoverable FreeSurfer recon-all and fMRIPrep preprocessing failures (Step 7; Supplementary material) meaning Freesurfer and/or fMRIPrep could not run successfully for these subjects/sessions. At the final QC stage, 28 sessions with excessive head motion (>30% censored frames or <5 minutes of low-motion data) or outlying Euler numbers were removed, yielding the final analysis-ready sample of 2,086 rs-fMRI sessions from 935 participants reported in Table 4.

The resulting sample spans the diagnostic spectrum captured by ADNI (cognitively unimpaired, MCI, and AD/dementia) and all three scanner manufacturers (Siemens, GE, and Philips) represented across ADNI GO/2/3 sites. The distribution and summary statistics of imaging-to-clinical-visit time gaps, as well as an overview of all longitudinal sessions, are visualized in Figure 4.

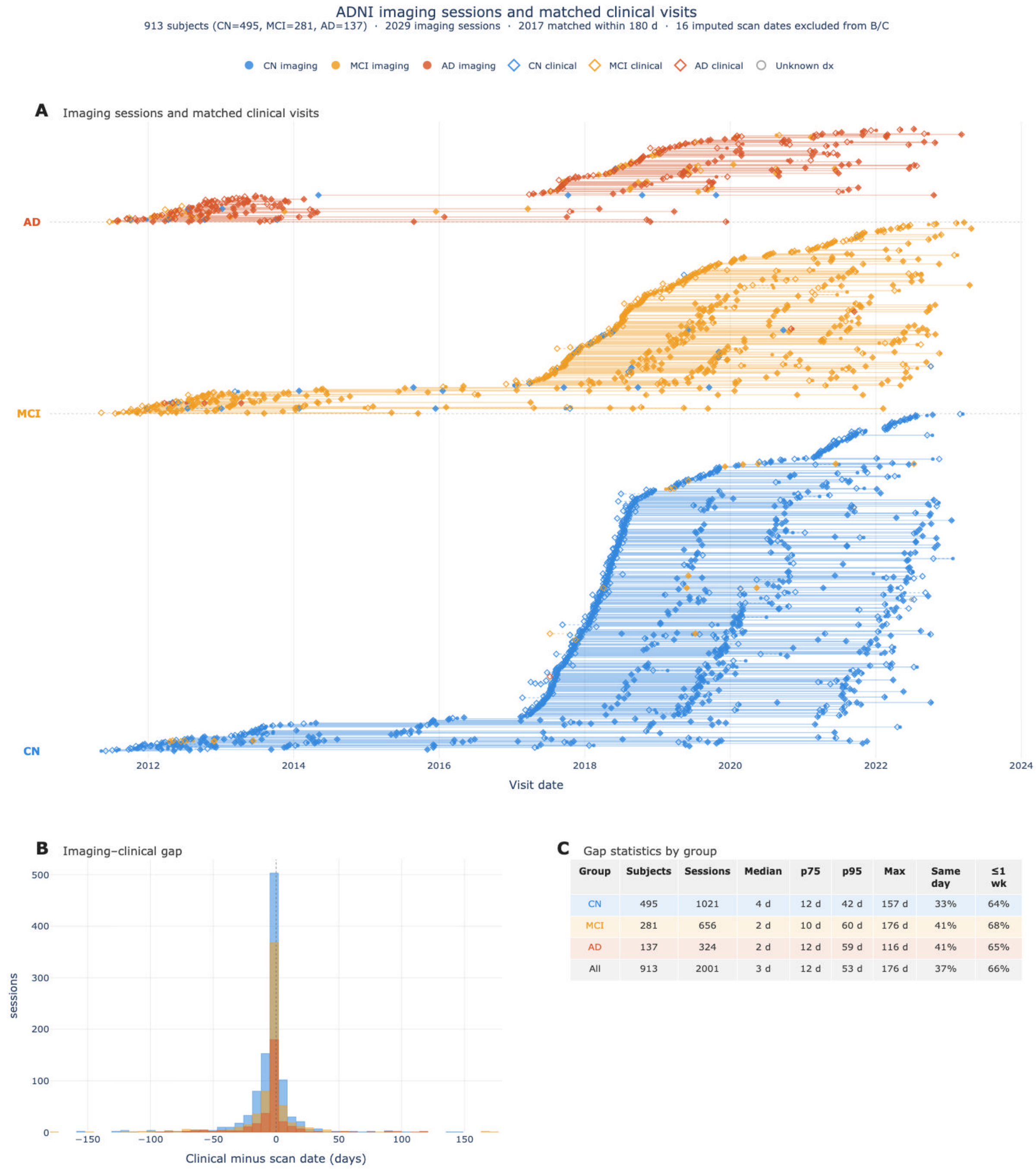

| Group | Subjects | Sessions | Median | p75 | p95 | Max | Same day | ≤1 wk |
|---|---|---|---|---|---|---|---|---|
| CN | 495 | 1021 | 4 d | 12 d | 42 d | 157 d | 33% | 64% |
| MCI | 281 | 656 | 2 d | 10 d | 60 d | 176 d | 41% | 68% |
| AD | 137 | 324 | 2 d | 12 d | 59 d | 116 d | 41% | 65% |
| All | 913 | 2001 | 3 d | 12 d | 53 d | 176 d | 37% | 66% |

**Figure 4. Imaging sessions and matched clinical assessments in the final ADNI rs-fMRI sample. A)** Visit timelines for all 913 subjects in the final sample, grouped by *final* diagnostic group (CN, MCI, AD). Filled circles denote imaging sessions, colored by diagnosis recorded at that visit; open diamonds denote clinical assessments; dotted segments connect a scan to its matched assessment and are therefore proportional to the interval between them. Solid horizontal lines link successive scans and are colored by final diagnosis. Subjects entering at later study phases appear as a staircase indicating a large temporal gap. **B)** Distribution of the signed interval between matched visits by diagnostic group. **C)** Summary of the absolute clinical-to-scan interval by diagnostic group. Panels B and C exclude 16 sessions whose scan dates were imputed and 12 sessions with no clinical visit within the 180-day matching window.

## 4 Discussion

Resting-state fMRI provides a non-invasive window into large-scale functional brain organization and has shown promise for detecting early network-level changes associated with cognitive decline and Alzheimer's disease[28,29]. Furthermore, ADNI is a richly deep and unique dataset whose impact continues to grow. Even so, ADNI rs-fMRI data have remained substantially underutilized relative to structural and PET imaging, largely due to the complexity of data curation, preprocessing, and quality control in a heterogeneous, multisite dataset.

In this work, we describe a data curation pipeline that addresses these challenges by providing a framework for transforming raw ADNI rs-fMRI data into an analysis-ready dataset linked to clinical assessments. The pipeline integrates clinical-imaging data alignment, standardized preprocessing, and multi-level quality control, with decisions at each stage documented and automated to support reproducibility. The pipeline lowers the technical barriers to using ADNI functional MRI data at scale. This is particularly important as research employing rs-fMRI increasingly moves toward large cohorts suitable for biomarker discovery, longitudinal modeling, and machine learning. This work enables researchers to leverage ADNI as a large longitudinal rs-fMRI resource for studying aging and AD. The resulting dataset facilitates the study of functional connectivity changes across disease stages, evaluation of fMRI-based prognostic markers, and integration of functional measures with the structural, PET, and fluid biomarkers already widely used in ADNI-based research.

Several limitations of the existing pipeline are worth noting. First, the pipeline relies on external tools, including Clinica, MRIQC, and fMRIPrep, and therefore inherits any limitations, issues, or design decisions associated with those tools. For instance, Clinica, which performs the essential step of conversion of ADNI data to BIDS format, does not currently support ADNI-4 data; accordingly, our pipeline currently covers through ADNI-3 only and will require corresponding upstream support to incorporate ADNI-4. Additionally, some of Clinica's data selection rules are not fully configurable by users, and its limited version control may inhibit full reproducibility. A related issue concerns the T1-weighted images available for a subset of ADNI sessions. In some cases, the only T1 image ADNI provides for a given session has already undergone processing (e.g., gradient-nonlinearity correction or intensity normalization) rather than representing the raw acquisition, and this processing history is not consistently documented in the accompanying metadata. Because Clinica selects among the T1 series that ADNI makes available, it cannot always distinguish a processed image from a raw one when ADNI itself does not flag the distinction. This highlights a need for more complete and standardized metadata and provenance tracking across the ADNI archive.

Several additional limitations arise from our own design decisions. For instance, while our quality control heuristics applied at Step 5 reflect expert judgment regarding plausible scan acquisition parameter ranges, the MRIQC IQM thresholds applied at Step 6 are based on an automated outlier detection technique using transformation to a normal distribution; these thresholds are not theoretically motivated, have not been validated against the stability of derived connectivity estimates, and depend on the quality of the normality transformation. IQMs may also benefit from site-specific thresholding, whereas we applied a universal threshold. Additionally, the 180-day window used to match clinical assessments to imaging sessions is generous relative to the observed distribution of intervals, and analyses treating cognitive scores as concurrent with imaging may warrant a more conservative threshold. Importantly, however, the pipeline allows investigators to adjust QC thresholds, window length, and other design decisions to suit the requirements of a given analysis and to assess the sensitivity of their results to these choices.

Future work will focus on extending this resource in ways that support the broader goals of aging and AD research. A priority is incorporating ADNI-4 data as upstream tool support becomes available, so that the pipeline continues to grow alongside the ADNI archive itself. We also plan to pursue harmonization strategies that reduce site- and scanner-related variability, making it easier to pool this resource with other multicenter datasets. Finally, we see promise in using large, curated resources like this one to inform individualized brain mapping, for instance by contributing to Bayesian priors that improve the precision of person-specific estimates of brain structure and function[30,31]. Together, these directions aim to make ADNI's functional imaging data an increasingly powerful resource for understanding aging and Alzheimer's disease.

In summary, we present a complete and open pipeline for transforming ADNI resting-state fMRI data into an analysis-ready resource. Applied to the full ADNI archive through ADNI-3, the pipeline yielded 935 participants and 2,086 sessions with curated and processed rs-fMRI data, verified scanner parameters and quality controls, and clinical assessments matched to each imaging session. A key strength of this work is the transparency and flexibility of the overall workflow: curation, preprocessing, and quality control decisions are documented and reproducible, while thresholds and other design choices can be adjusted to meet the needs of a given analysis. By facilitating more widespread and large-scale use of the ADNI resting-state fMRI data, this resource will support studies of functional brain changes and biomarker development in aging and Alzheimer's disease. More broadly, this work provides a practical framework for bringing legacy neuroimaging datasets into contemporary, reproducible research workflows to accelerate reuse of existing neuroimaging resources.

## 5 Code Availability Statement

All code required to reproduce the complete ADNI resting-state fMRI preprocessing workflow is openly available in the accompanying GitHub repository (https://github.com/STATMINDlab/ADNI_fMRI_protocol). All workflows and scripts are version controlled. A static, citable snapshot of the repository (including the configuration files, QC code, and documentation) will be archived on Zenodo, and a DOI will be provided upon publication. This ensures that any researcher with ADNI data access can regenerate every derivative, QC table, and processing decision described in this manuscript. Full details of the repository structure and file contents are provided in the Supplementary Material.

## Disclosures

The authors have no disclosures to report.

## Consent Statement

All subjects provided informed consent in the original ADNI study protocol.

## Author Contributions

S.R. downloaded and processed the data, wrote the code and tests, developed the pipeline, prepared the figures, wrote the original draft of the manuscript, and edited the manuscript. Z.Z. wrote and tested the code, processed data, and prepared selected figures. A.A.K. supervised data processing, tested the code and reviewed the manuscript. R.C.W supervised data processing and reviewed the code and manuscript. D.H. reviewed the manuscript. V.K. downloaded the data, supervised data processing, developed the pipeline, and reviewed the code and manuscript. A.F.M conceptualized the project, provided supervision, obtained funding, and edited the manuscript.

## Acknowledgements

Data collection and sharing for this project was funded by the Alzheimer's Disease Neuroimaging Initiative (ADNI) (National Institutes of Health Grant U01 AG024904) and DOD ADNI (Department of Defense award number W81XWH-12-2-0012). ADNI is funded by the National Institute on Aging, the National Institute of Biomedical Imaging and Bioengineering, and through generous contributions from the following: AbbVie, Alzheimer's Association; Alzheimer's Drug Discovery Foundation; Araclon Biotech; BioClinica, Inc.; Biogen; Bristol-Myers Squibb Company; CereSpir, Inc.; Cogstate; Eisai Inc.; Elan Pharmaceuticals, Inc.; Eli Lilly and Company; EuroImmun; F. Hoffmann-La Roche Ltd and its affiliated company Genentech, Inc.; Fujirebio; GE Healthcare; IXICO Ltd.; Janssen Alzheimer Immunotherapy Research & Development, LLC.; Johnson & Johnson Pharmaceutical Research & Development LLC.; Lumosity; Lundbeck; Merck & Co., Inc.; Meso Scale Diagnostics, LLC.; NeuroRx Research; Neurotrack Technologies; Novartis Pharmaceuticals Corporation; Pfizer Inc.; Piramal Imaging; Servier; Takeda Pharmaceutical Company; and Transition Therapeutics. The Canadian Institutes of Health Research is providing funds to support ADNI clinical sites in Canada. Private sector contributions are facilitated by the Foundation for the National Institutes of Health. The grantee organization is the Northern California Institute for Research and Education, and the study is coordinated by the Alzheimer's Therapeutic Research Institute at the University of Southern California.

## Sources of Funding

This work was supported by grants from the National Institutes of Health/National Institute on Aging R01AG083919 and K01AG073578.

# Supplemental Materials

## Supplemental Methods

***PubMed Search Strategy***

The publication count (Main Text Figure 1) was obtained from a PubMed search using the terms "Alzheimer's Disease Neuroimaging Initiative" AND ("fMRI" OR "functional MRI" OR "rsfMRI" OR "resting state"), restricted to records dated through June 2026. Once the list of papers was compiled from PubMed, a research assistant reviewed all papers to confirm a) that the paper used ADNI rs-fMRI data and b) the reported ADNI sample size. The exact database of papers (including metadata (i.e., journal, publication date, and author) and their reported sample sizes is included as a supplemental file (ADNI_pubmed_rsfmri_papers.csv).

*FMRIPrep preprocessing methods*

*Anatomical preprocessing*

The T1 image for each subject is corrected for intensity non-uniformity (INU) with N4BiasFieldCorrection[1], distributed with ANTs 2.6.2[2], and used as T1 reference throughout the workflow. The T1 reference is then skull-stripped with a Nipype implementation of the antsBrainExtraction.sh workflow (from ANTs), using OASIS30ANTs as target template. Brain tissue segmentation of cerebrospinal fluid (CSF), white-matter (WM) and gray-matter (GM) is performed on the brain-extracted T1 using FSL fast[3]. Brain surfaces are reconstructed using recon-all (FreeSurfer 7.3.2)[4], and the brain mask estimated previously is refined with a custom variation of the method to reconcile ANTs-derived and FreeSurfer-derived segmentations of the cortical gray-matter of Mindboggle[5]. A FLAIR image is used to improve pial surface refinement. Brain surfaces are reconstructed using recon-all (FreeSurfer 7.3.2)[4], and the brain mask estimated previously was refined with a custom variation of the method to reconcile ANTs-derived and FreeSurfer-derived segmentations of the cortical gray-matter of Mindboggle[5]. Volume-based spatial normalization to two standard spaces (MNI152NLin6Asym, MNI152NLin2009cAsym) is performed through nonlinear registration with antsRegistration (ANTs 2.6.2), using brain-extracted versions of both T1 reference and the T1 template. The following templates were selected for spatial normalization and accessed with TemplateFlow[6] (25.0.4): FSL's MNI ICBM 152 non-linear 6th Generation Asymmetric Average Brain Stereotaxic Registration Model[7] [TemplateFlow ID: MNI152NLin6Asym], ICBM 152 Nonlinear Asymmetrical template version 2009c[8] [TemplateFlow ID: MNI152NLin2009cAsym].

*Functional preprocessing*

First, a BOLD reference volume is generated for use in head motion correction. Head-motion parameters with respect to the BOLD reference (transformation matrices, and six corresponding rotation and translation parameters) are estimated before spatiotemporal filtering using FSL mcflirt[9]. The BOLD reference is then co-registered to the T1 reference using bbregister (FreeSurfer), which implements boundary-based registration[10]. Co-registration is configured with six degrees of freedom.

Several confounding time series are calculated based on the preprocessed BOLD: framewise displacement (FD), DVARS, and three region-wise global signals. FD is computed using two formulations: the Power formulation (absolute sum of relative motions)[11] and the Jenkinson formulation (relative root-mean-square displacement between affines)[12]. FD and DVARS are calculated for each functional run, both using their implementations in Nipype (following the definitions by Power[11]). The three global signals are extracted within the CSF, the WM, and the whole-brain masks.

Additionally, a set of physiological regressors are extracted to allow for component-based noise correction (CompCor[13]). Principal components are estimated after high-pass filtering the preprocessed BOLD time series (using a discrete cosine filter with 128s cut-off) for the two CompCor variants: temporal (tCompCor) and anatomical (aCompCor). tCompCor components were calculated from the top 2% variable voxels within the brain mask. For aCompCor, three probabilistic masks (CSF, WM and combined CSF+WM) are generated in anatomical space. The implementation in fMRIprep differs from that of Behzadi et al.[13] in that, instead of eroding the masks by 2 pixels on BOLD space, a mask of pixels that likely contain a volume fraction of GM is removed from the WM and CSF masks. This mask is obtained by dilating a GM mask extracted from the FreeSurfer's aseg segmentation, and it ensures components are not extracted from voxels containing a pre-set minimal fraction of GM. Finally, these masks are resampled into BOLD space and binarized by thresholding at 0.99 (as in the original implementation). Components are also calculated separately within the WM and CSF masks. For each CompCor decomposition, the k components with the largest singular values are retained, such that the retained components' time series are sufficient to explain 50 percent of the variance across the nuisance mask. The remaining components are dropped from consideration.

The head-motion estimates calculated in the correction step are also placed within the corresponding confounds file. The confound time series derived from head motion estimates and global signals are expanded with the inclusion of temporal derivatives and quadratic terms for each[14]. Frames that exceed a threshold of 0.5 mm FD or 1.5 standardized DVARS are annotated as motion outliers. Additional nuisance timeseries are calculated by means of principal components analysis of the signal found within a thin band (crown) of voxels around the edge of the brain, as proposed by Patriat et al.[15].

The BOLD time series are resampled onto the following surfaces (Freesurfer reconstruction nomenclature): fsnative, fsaverage5, fsLR. The BOLD time series are resampled from T1 native (subject-specific) volumetric space into surface native (subject-specific) space, and then the surface native mesh is resampled onto the left/right-symmetric group-level template "fsLR" using the Connectome Workbench[16]. Grayordinates files [16] containing 91k cortical and subcortical coordinates are also generated with surface data transformed directly to fsLR space and subcortical data transformed to 2 mm resolution MNI152NLin6Asym space. All resamplings can be performed with a single interpolation step by composing all the pertinent transformations (i.e., head-motion transform matrices, susceptibility distortion correction when available, and co-registrations to anatomical and output spaces). Gridded (volumetric) resamplings are performed using nitransforms, configured with cubic B-spline interpolation. Non-gridded (surface) resamplings are performed using mri_vol2surf (Freesurfer).

*ADNI specific fMRIPrep preprocessing errors*

Missing BOLD file errors occur when subjects have multiple session directories but only a subset contain functional data (e.g., ses-M000/, ses-M006/, ses-M012/, ses-M048/, ses-M096/, with BOLD present only in later sessions). To prevent fMRIPrep failures in these cases, the pipeline explicitly inspects session directories prior to submission and dynamically restricts fMRIPrep execution to sessions that contain valid BOLD files. As a result, fMRIPrep jobs are only launched for subject–session pairs that meet minimum data requirements. This logic is implemented in the s7_fmriprep/fMRIPrep_rerun_array.slurm script and is applied uniformly across all subjects.

TemplateFlow binding and memory-related errors affect only three subjects in our data download. These errors are resolved by adjusting resource allocation and ensuring that the TemplateFlow directory is correctly mounted inside the container. The provided fMRIPrep execution scripts automatically download and bind the TemplateFlow directory, allowing these cases to be successfully reprocessed without further intervention.

FreeSurfer recon-all failures reflect unrecoverable errors in cortical surface reconstruction, most likely due to low-quality anatomical data. Subjects with these failures are excluded from downstream analyses, as these errors cannot be reliably corrected.

Syn-sdc failures are caused by missing or invalid PhaseEncodingDirection metadata in the BOLD JSON files, which is common for certain ADNI acquisitions (primarily from Philips scanners). Because PhaseEncodingDirection is required for fieldmap-less distortion correction, these runs fail when syn-sdc is enabled. Our error classification automatically detects this condition (find_syn_pedir_failures.py) and re-runs affected subjects without the –use-syn-sdc flag using s7_fmriprep/fMRIPrep_rerun_synflag_array.slurm. Users may choose to exclude these subjects due to these differences in preprocessing; however, we retain them in the final dataset, as ADNI already contains unavoidable heterogeneity in acquisition parameters and available anatomical images (e.g., presence or absence of T2 scans).

## Supplemental Figures

The figures below show representative sessions flagged by the automated IQM outlier-detection procedure (Section 2.6) for each MRIQC metric. Note: examples below are currently drawn from a single selection pass; paired random and near-threshold examples for every metric. If only one image is shown, it means the near-threshold and at random were chosen for the same subject, or there are too few outliers to show more than one example subject. For the T1w IQMs, the Freesurfer recon-all surface reconstruction image is shown where the pial surface (blue) and cortical surface (red) lines are shown. For the BOLD IQMs, carpet plots of the time-series are shown.

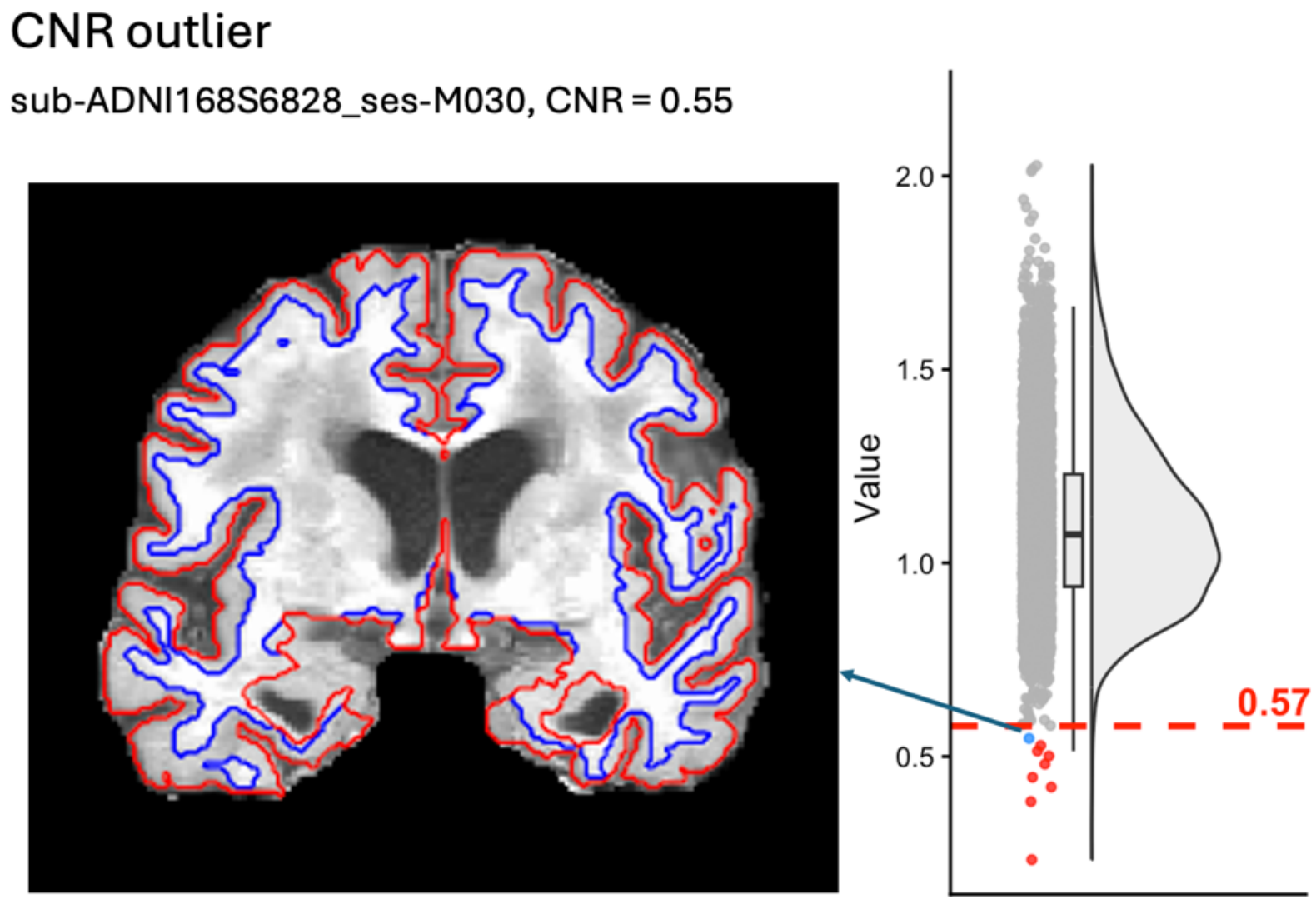

**Supplementary Figure 1.** Example T1w session flagged as a Contrast-to-Noise Ratio (CNR) outlier. The figure on the left is the blue dot near the outlier threshold.

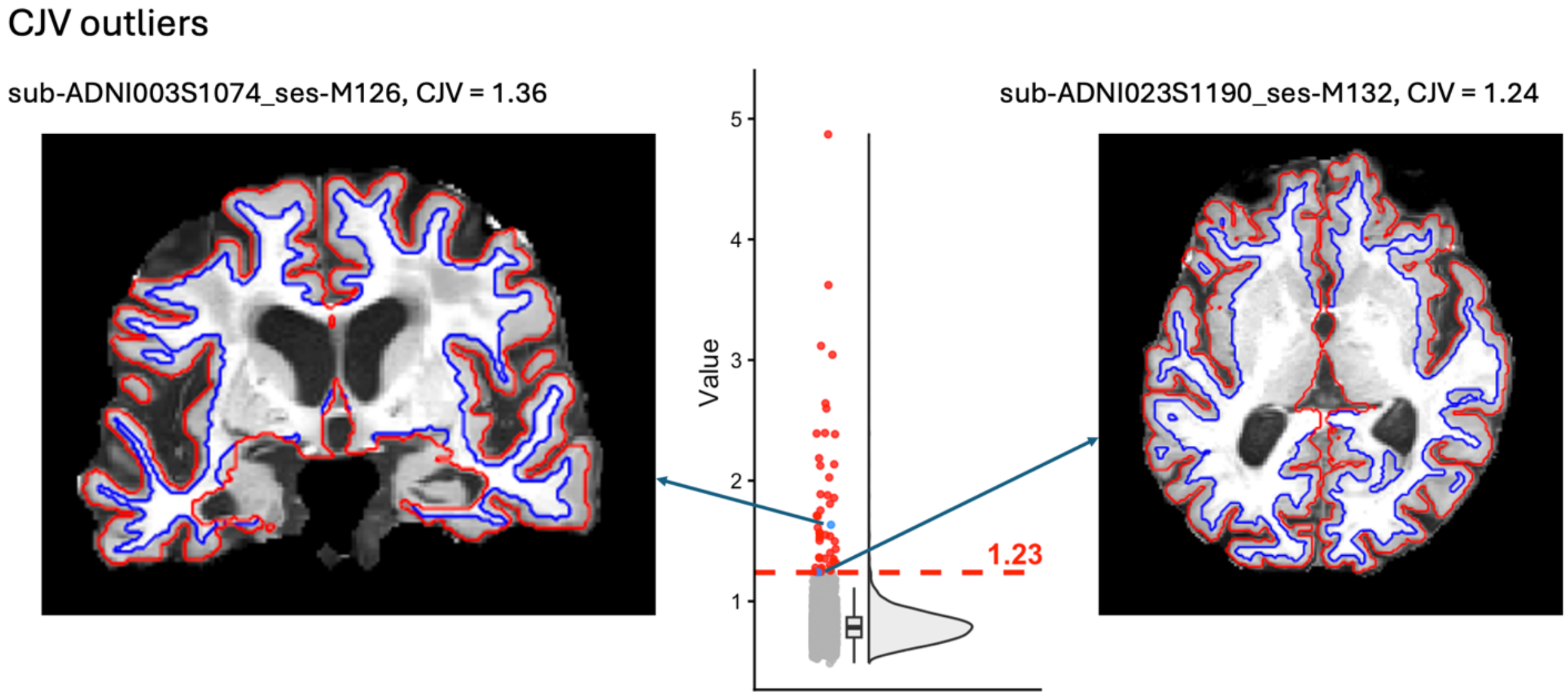

**Supplementary Figure 2.** Example T1w sessions flagged as Coefficient of Joint Variation (CJV) outliers.

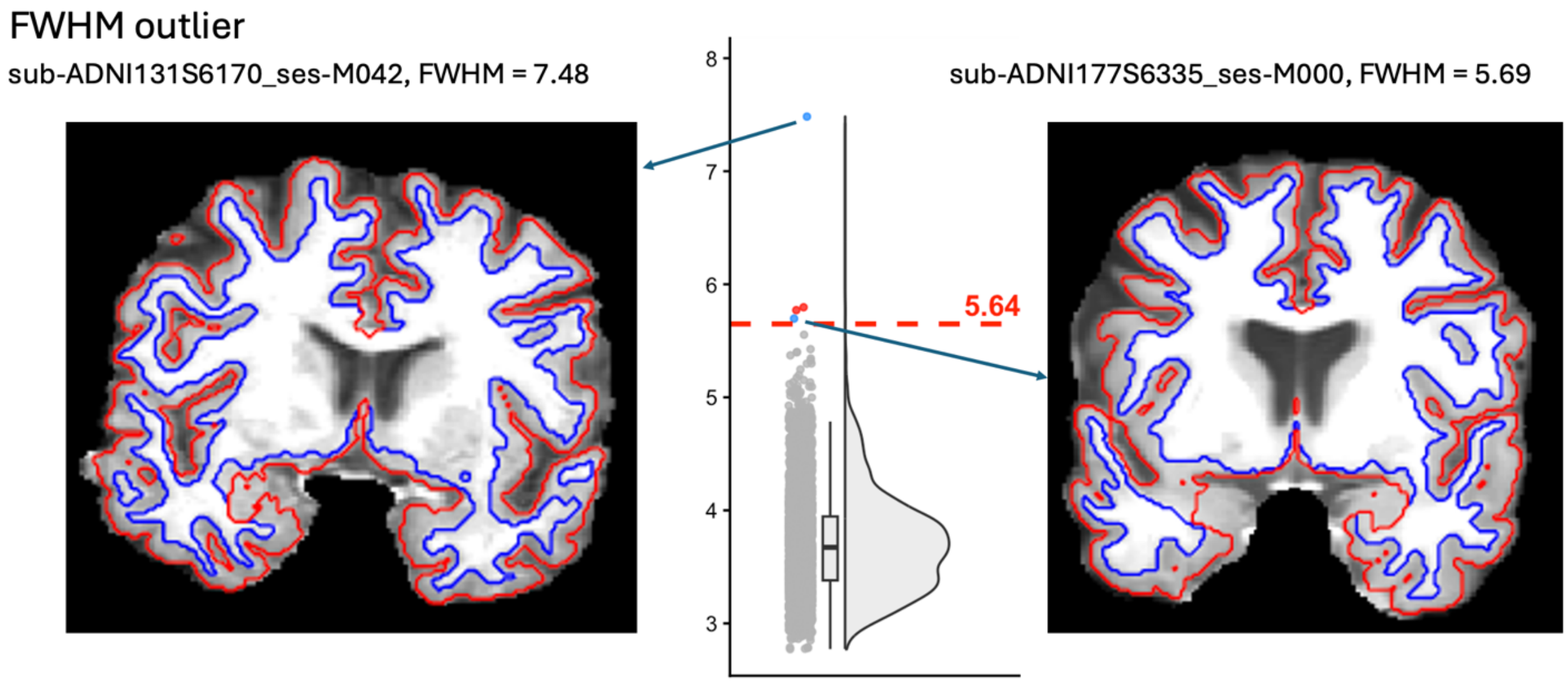

**Supplementary Figure 3.** Example T1w sessions flagged as a Full Width at Half Maximum (FWHM) outlier.

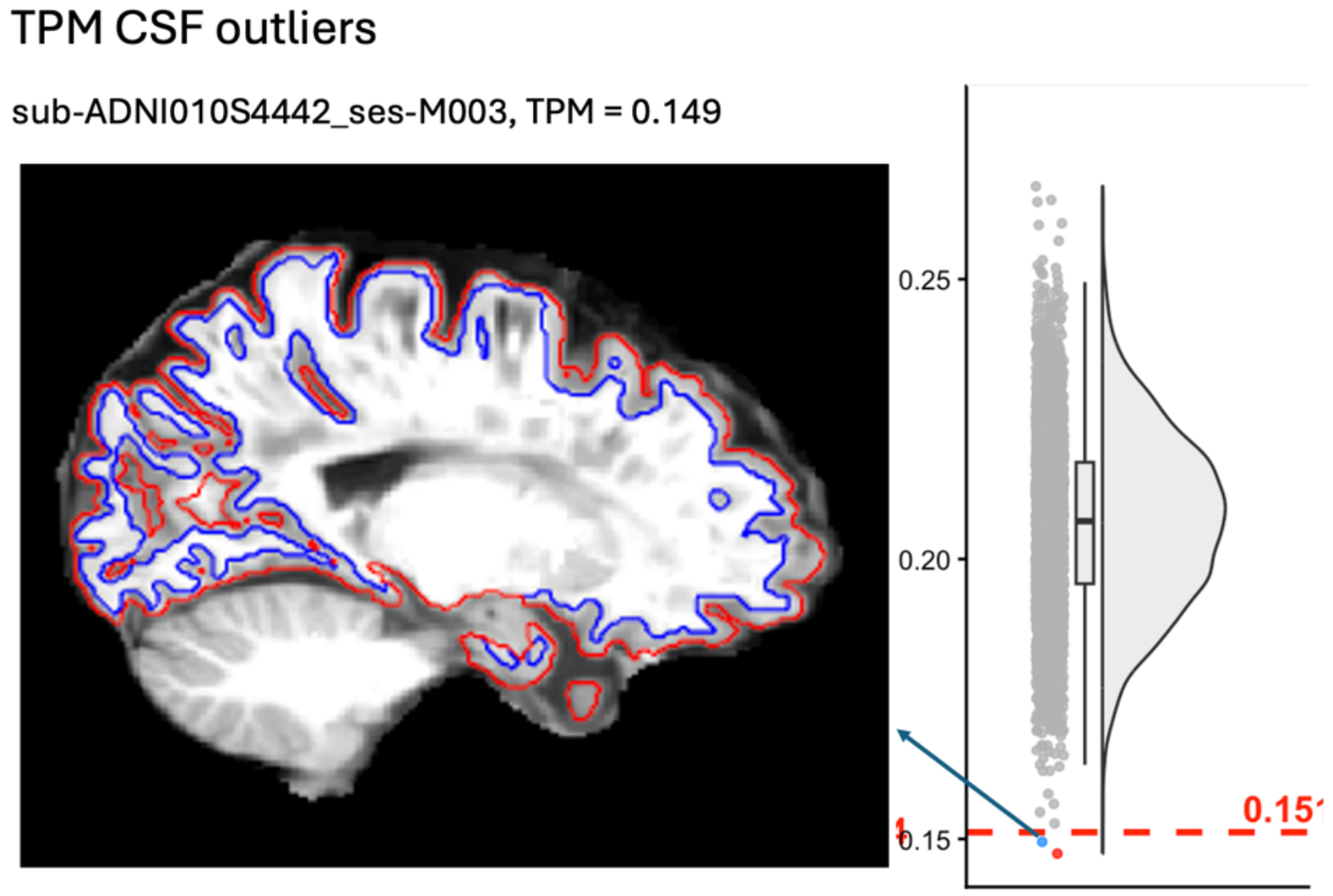

**Supplementary Figure 4.** Example T1w session flagged as a Tissue Probability Map (TPM) overlap outlier.

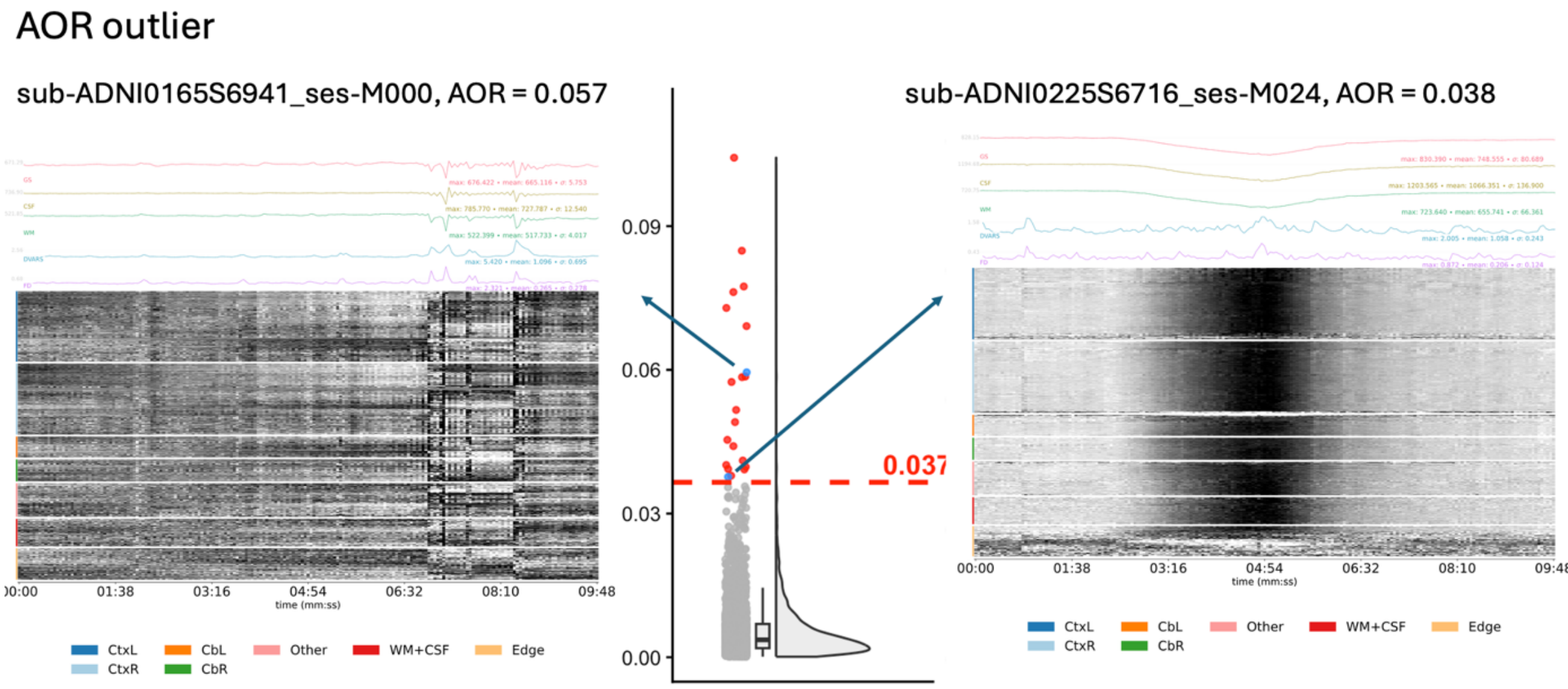

**Supplementary Figure 5.** Example BOLD session flagged as an AFNI outlier ratio (AOR) outlier, shown as a carpet plot with DVARS and FD traces.

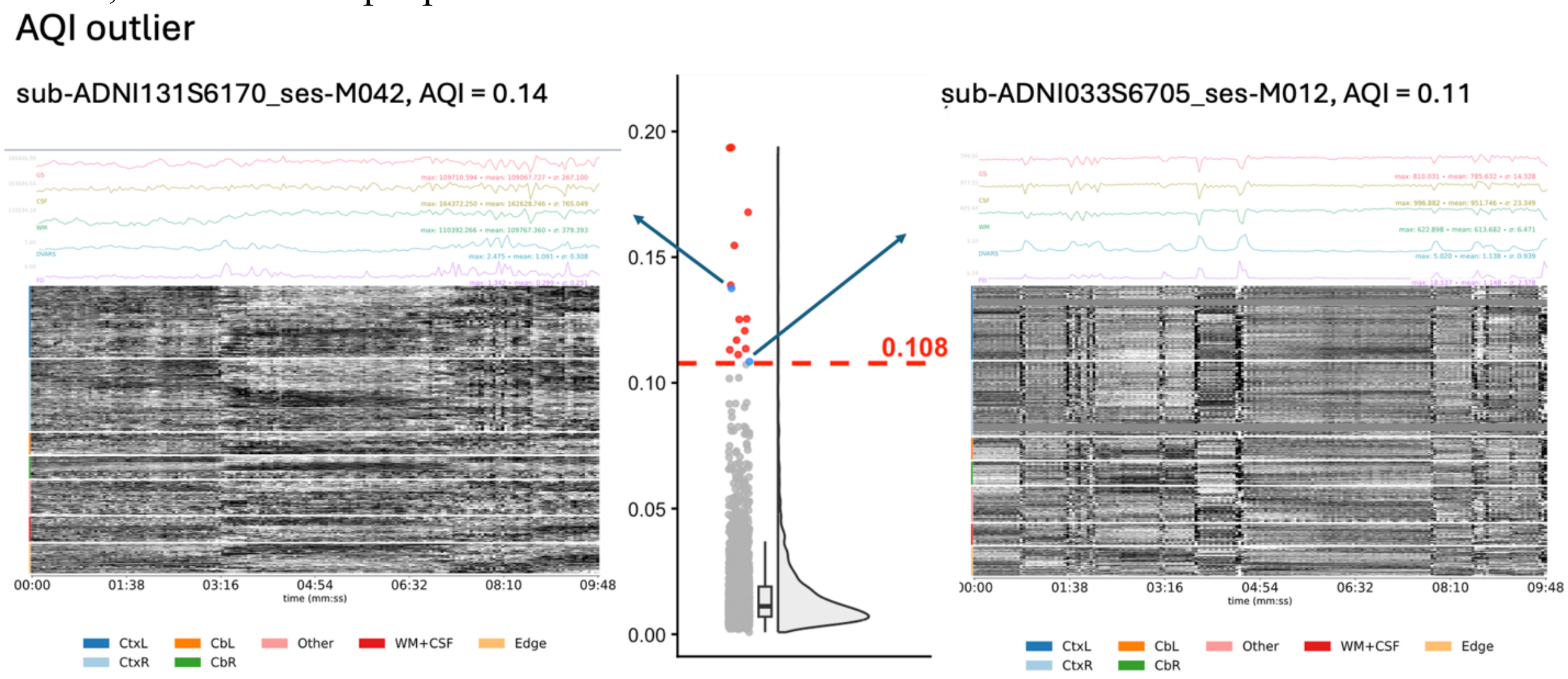

**Supplementary Figure 6.** Example BOLD session flagged as an AFNI quality index (AQI) outlier, shown as a carpet plot with DVARS and FD traces.

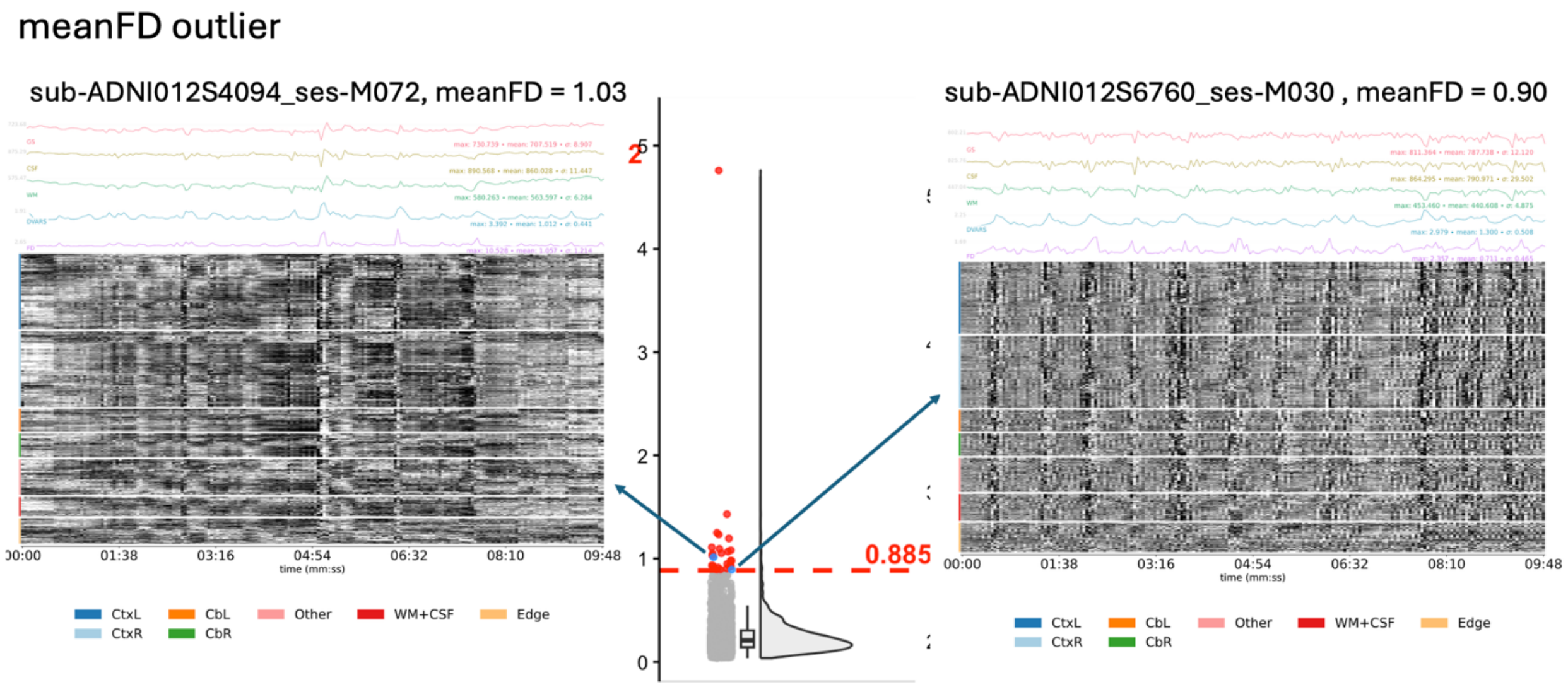

**Supplementary Figure 7.** Example BOLD sessions flagged as a mean Framewise Displacement (meanFD) outlier, shown as a carpet plot with DVARS and FD traces.

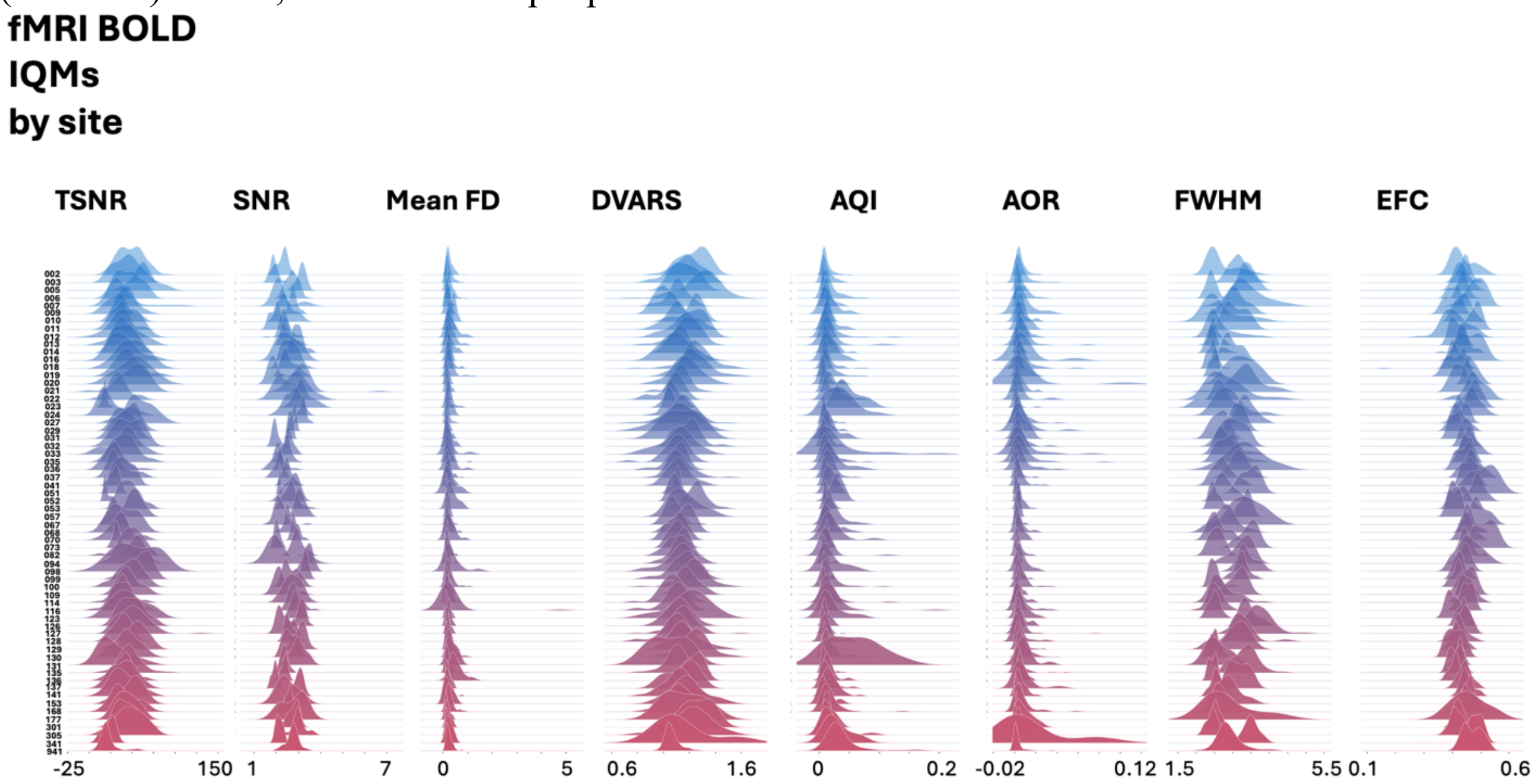

**Supplementary Figure 8.** Distribution of BOLD image quality metrics (IQMs) by site.

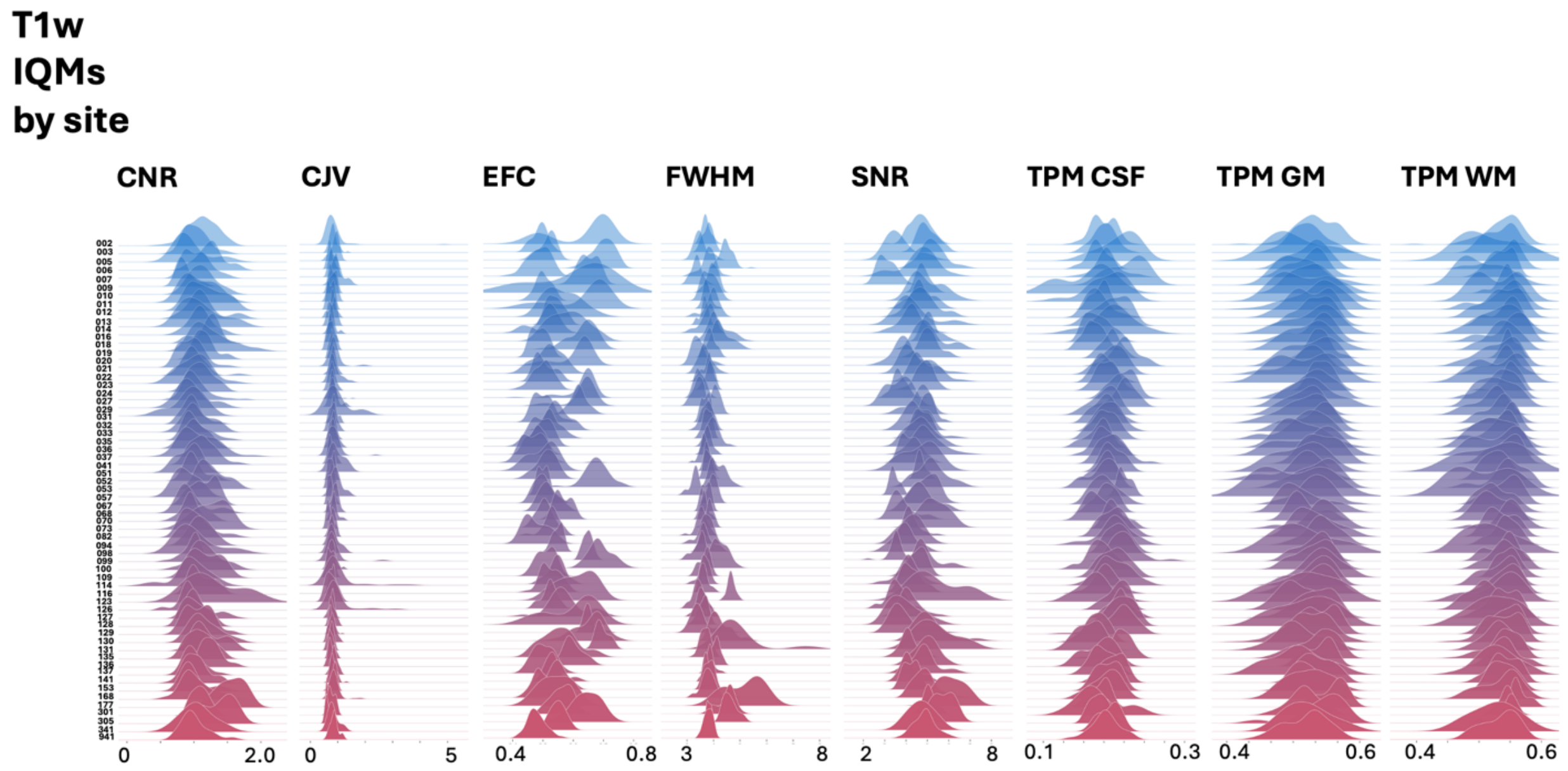

**Supplementary Figure 9.** Distribution of T1w image quality metrics (IQMs) by site.

## Supplemental References